\documentclass[twocolumn,showpacs,preprintnumbers,amsmath,amssymb,aps,prb]{revtex4-1}

\usepackage{graphicx}
\usepackage{dcolumn}
\usepackage{bm}
\usepackage{natbib}
\usepackage{amssymb}
\usepackage{hyperref}

\begin{document}

\title{Background Free Measurement of the Spectra of Low Energy Electrons Emitted as a Result of Auger Transitions in Metals }

\author{
S.F. Mukherjee,  K. Shastry,  A.H. Weiss}

\affiliation{
Department of Physics, University of Texas at Arlington, 
Arlington, TX, USA,76019
}


\begin{abstract} 
Positron annihilation induced Auger electron spectroscopy was used to obtain Cu and Au  Auger 
spectra that are free of primary beam induced background by impinging the positrons at energy
 below the secondary electron emission threshold. The removal of the core electron via annihilation
 in the PAES process resulted in the elimination of post-collision affects.  The spectrum indicates 
that there is an intense low energy tail (LET) associated with the Auger peak that extends all the
 way to $0$ eV. The LET has been interpreted as being due to processes in which the filling of the
 core by a valence electron results in the ejection of two or more valence electrons which share
 the energy of the otherwise Auger electron

\end{abstract}


\maketitle

\section{Introduction}
\label{Introduction}
Electron and X-ray induced Auger electron spectroscopy (EAES and XAES) are extremely sensitive 
to the composition and chemistry of the surface \cite{ramaker:1614, PhysRevLett.94.038302, PhysRevLett.93.206802}
 owing to the short inelastic mean free path of electrons ($5\text{\AA}-20\text{\AA}$) emitted in the Auger energy range
 ($\sim 20$- $2000$ eV) \cite{tougaard1988quantitative, seah1972quantitative}.  Both these techniques 
use an incident beam energy greater than the ionization energy of the core electron ($E_{C}$) that
 is removed in order to initiate the Auger process. The energy of the emitted electron as a result
 of filling up of the initial hole is less than $E_C$ due to the energy associated with the
 two hole final state of the Auger process. Consequently, the Auger peaks in EAES and XAES are
 superimposed upon a background composed of an electron cascade
 \cite {Seah1969132, Sickafus1977SecE2, SickafusSecE1} caused by the inelastic scattering of 
the incident electron beam or photo excited valence electrons. Such background tend to be many
 times the intensity of the Auger peak itself.  In both the cases of EAES and XAES the background
 is considered extrinsic to the Auger process and is typically subtracted
 \cite{seah1996quantitative, contini1989restoration, ramaker1979extracting}
 to reveal the true Auger spectrum.

In this paper, we present full range spectrum of the electrons emitted following the creation 
of core hole using positron annihilation induced Auger electron spectroscopy (PAES) (as shown
 in Fig.~1a). Recently, it has been shown that it is possible to efficiently trap positrons directly in
 a surface state  and excite Auger electrons by using a low positron beam energy ($\sim 1.5$ eV) which
 is well below the threshold for secondary electron emission \cite{Mukherjee2010AMS}. Here, we
 have used this effect to measure the first completely background free Auger spectrum down to $0$ eV.
  The surface state trapped positrons tunnel into the core region and annihilate with the electrons 
leading to the Auger process. The localization of the positrons in the surface state ensures that 
almost all of the excitations that result in Auger transition occur in the topmost atomic layer
 \cite{PhysRevB.41.3928}. This is in contrast to conventional Auger spectroscopy (XAES or EAES)
 where, due to the penetrating  nature of the incident beam, Auger electrons are excited at depths
 far in excess of the electron inelastic mean free path. Such spectra are further complicated by post
 collision interaction (PCI) effect which refers to the interaction of the excited system
 (solid with a core hole) with the excited core electron. Such effect leads to plasmon loss features in 
X-ray photoelectron spectra and a consequent plasmon gain  feature in the Auger spectra. This 
effect is absent in PAES as the Auger process is initiated by annihilation of the core electron. The
 low energy signals from conventional AES also contain the beam induced background which is
 typically many times more intense than the signal due to the Auger process. In clear contrast,
 a $1.5$ eV positron induced PAES spectrum contains contributions that result directly or indirectly
 form Auger transitions in the top-most atomic layer.

\begin{figure} 
\begin{center} 
\includegraphics[width=\hsize,width=8.cm]{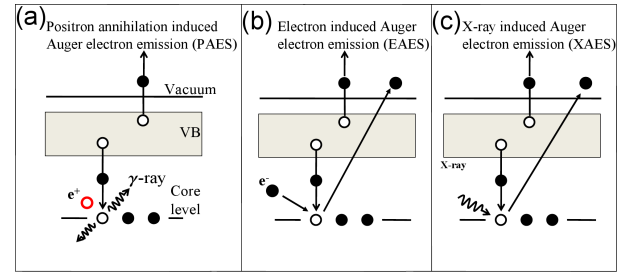} 
\end{center} 
\caption{(Color online)
Energy band diagram showing mechanism for (a) PAES (b) EAES and (c) XAES. In PAES, the core hole is
 created by matter-antimatter annihilation and hence it is possible to get Auger emission with incident
 positron energy $E_{p}\to 0$ eV. As opposed to EAES and XAES, the post collision interaction (PCI) 
effects are absent in PAES since the core electron is annihilated.
} 
\label{fig1} 
\end{figure}

Based on these considerations,  PAES spectra can be understood and analyzed based on just
 two factors- the usual Auger peak (C-VV transition) and the intrinsic loss(C-VVV transition)\cite{ramaker:1614}
 [C=Core, V= Valence]. The intrinsic loss 
associated with creation of a core hole has mostly been studied in context of the asymmetricity
 in the photopeak and presence of intrinsic plasmons in X-ray photoelectron spectroscopy (XPS) 
and is refered to as Mahan-Nozieres-DeDominicis effect \cite{ MahanExcitons, NozieresSingularities, 
DoniachSingularitiesXPS,  tougaard1988quantitative, HufnerXPS}. The intrinsic loss reflects as
 a tail on the low energy side of the core peak and is assumed to extend to $\sim50 eV$ below the
 peak \cite{tougaard1988quantitative}. The distinction between intrinsic and extrinsic electrons is based 
on the observation that the latter were related to electron transport in the solid while the former were due
 to creation of the core hole. Since the creation of core hole and the associated Auger process 
are essentially a single step process \cite{JensenLineNarrowing} similar assymetricities should be 
present in Auger peaks as well. Experimental verification of this has been attempted solely by
 Auger-photoelectron coincidence spectroscopy (APECS) where a clear distinction between
 extrinsic and intrinsic electrons cannot be made\cite{JensenAPECS, ThurgateIntrinsic, VanRiessenDPE}. Hence an 
exact quantitative and qualitative prediction over the whole energy range has been missing. 
In PAES the core hole creation probability is independent of the incident positron energy
 which allows us to set the incident positron energy ($E_{p}=1.5$ eV) well below the secondary
 electron creation threshold. The resultant spectra show a low energy tail (LET) \cite{JensenAPECS}
 with intensity ($I_{LET}$)$\sim3.92(3.43)$ times that of the Auger peak intensity ($I_{Auger}$) for Cu(Au) at
 energies below the Auger energy and extending to $0$ eV.  The LET has been interpreted as arising
 mostly from intrinsic loss associated with the creation of the core hole. The spectral weight of intrinsic
 part of the LET is $~2.39(3.0)$ times that of the Auger peak area which has been concluded as
 evidence of multi electron emission once the core hole is created. This process is analogous to the 
double Auger process in noble gases \cite{PhysRevLett.14.390} and similar to photon induced 
correlated electron emission from solids \cite{PhysRevLett.89.086402}. The results have been
 interpreted as signature of electron correlation in the valence band \cite{JensenAPECS, VanRiessenDPE}.
\section{Experimental}
\label{Experimental}The experiments were carried out in the time of flight -positron annihilation 
induced Auger electron spectrometer (tof-PAES) \cite{xie2002positron} which uses a magnetic
 bottle analyzer \cite{JPhysESciIns.16.313}. Positrons emitted via beta decay from a $4$ mCi $Na-22$
 source were moderated using a polycrystalline Tungsten (W). The upper limit on the positron beam 
energy, $E_P$, is given by the following equation
 \begin{equation}
E_{\mbox{p}}=\varphi_m^{+}+e(V_{m}-V_{s})
\end{equation}   
where $\varphi_m^{+}$ is the positron 
work function of the moderator($2.9$ eV for W), $V_{m}$  and $V_{s}$ are the bias on the moderator and
 sample with respect to the ground and $e$  is the electronic charge. The low energy 
positrons are guided to the sample, located $3.5$ m away, using $EXB$ fields. The whole 
spectrometer is housed in Helmholtz cage to cancel out the earth’s magnetic field. 
The energy of the outgoing electron is referred to the vacuum level.
\begin{figure} 
\begin{center} 
\includegraphics[width=\hsize,width=6.cm]{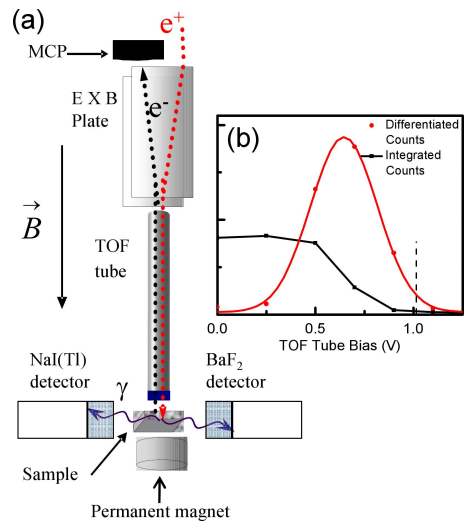} 
\end{center} 
\caption{(Color online)
Experimental setup for time of flight-positron annihilation induced Auger electron
 spectroscopy (tof-PAES)(a) Schematic of the spectrometer.(b)The incident positron 
beam energy used to obtain the Auger spectra. The dashed line($1$ eV) refers to the incident
 positron energy and $99\% $of positrons have energy less than this value ($e^+$ =positrons, $e^-$ =electrons, B= magnetic field)
} 
\label{fig2} 
\end{figure}

 The annihilation gamma rays are detected by BaF$_2$ and NaI detectors as shown in Fig.~2a. 
The outgoing electrons from the sample are parallelized using the divergent field of a permanent
 magnet. The electrons then fly down a retarding time of flight tube (tof tube) and are
 detected by a Micro channel plate (MCP). The MCP signal is used as the START while the
 delayed BaF$_2$ signal provides the STOP (reverse timing) of the Time to Amplitude convertor (TAC). 
The TOF-PAES spectra are obtained by histogramming the output of the TAC. A calibration
 procedures detailed in reference \cite{Mukherjee2010AMS} was used to determine the relation between
 the measured time of flight and the kinetic energy of the electrons leaving the sample.  To estimate
 the contribution of accidental coincidence to tof-PAES spectra, a second setup referred to as the 
triple coincidence setup was used which takes advantage of the fact the annihilating gamma rays
are emitted at an angle of $\sim 180^o$ with respect to each other. In this the STOP signal is the coincidence
 detection of the collinear $511$ keV gamma rays  by BaF$_2$ and NaI(Tl).  The triple coincidence set 
up, by requiring the detection of two almost antiparallel gamma rays, was designed to discriminate 
against events in which one or both of the annihilation gamma rays generate secondary electrons 
as a result of Compton scattering in the sample or surrounding chamber walls.  


The $4$ mT transport field used in the TOF spectrometer is particularly well suited for efficiently 
transporting the low energy positron (which was incident at $1.5$ eV at the sample) as well as the
 low energy electrons emitted from the sample.   A $-0.5$ volt sample bias was used to boost the
 kinetic energy of electrons emitted from the surface permitting measurements of electrons emitted 
from the surface all the way down to $0$ eV. The time of flight for a $0.5$ eV electron (corresponding to $0$ eV
 KE emission from the sample) was $\sim2 \mu s$ which was well within the measuring range of the spectrometer.


The incident beam profile (shown in Fig.~2(b))at $0$ eV sample bias was
 fitted with a Gaussian of $0.4$ eV FWHM 
and maximum at $0.65$ eV. $99$ \% of the positrons have energy 
less than  $1$ eV and this is referred to as the beam energy. This beam energy was
 used to obtain the Auger spectra of Cu and Au. During the measurements of the
 Auger spectra, the sample was biased at $-0.5$ V with respect to the TOF drift tube.

%

 An Au sample (a $99.985$ \% pure polycrystalline foil, $0.025$ mm thickness) was sputter 
cleaned every $12$ hours while a Cu(100) sample (a $99.9$ \% pure, $10$ mm diameter
 $\times$ $1$ mm thickness) was sputter cleaned followed by annealing at $740$ 
$^o$C every $12$ hours. The PAES spectrum was used to monitor the cleanliness 
of the sample and no significant contamination of the surface was found in the period 
between two sputtering times.
%

%


 The sensitivity of PAES to the low energy electrons ($\geq 0.5$ eV) is demonstrated in spectra
 shown in Fig.~3 which were obtained with an incident beam energy of $2$ eV.  The incident positron
 energy was changed by varying the moderator bias (Eq~1) while keeping the electric and 
magnetic fields between the sample and MCP same. The peaks in the timing spectra (Fig.~3)
 are the Auger peak and the positron sticking induced secondary electron peak \cite{Mukherjee2010AMS}. 
The lagging edge of the secondary electron  peak (corresponding to the longest flight times)  moves to 
shorter  flight time as the sample bias is increased from$ -0.5$ V to $-1$ V.   These results confirm the ability
 of the tof-PAES system to detect and measure the energy of electrons emitted from the sample down
 to sub-eV energies. To obtain the Auger spectra, the moderator voltage was changed back to get a 
$1$ eV beam incident on the sample at $0$V bias. All the Auger spectra were taken with this setting and
 sample bias $-0.5$V.

\begin{figure} 
\begin{center} 
\includegraphics[width=\hsize,width=8.cm]{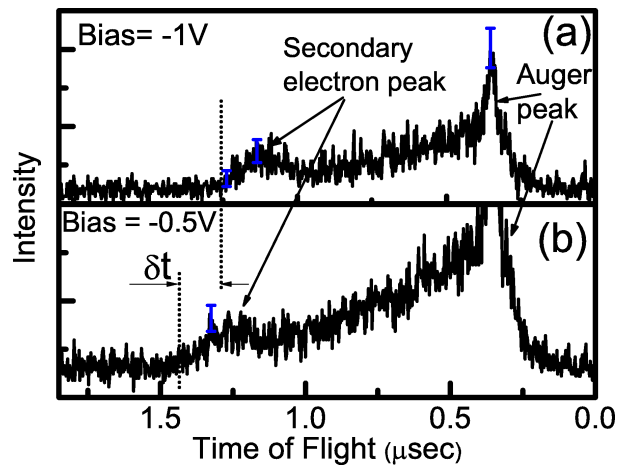} 
\end{center} 
\caption{
PAES spectra from Au obtained with a bias of $-1$V and $-0.5$ V.  The spectra contains sticking 
 induced secondary electron peak \cite{Mukherjee2010AMS} and Auger peak. The incident positron beam
 energy at $0$ V sample bias is $2$ eV.  The spectra is sitting on a flat  background due to 
 accidental coincidences. The vertical dotted line in both the panels shows the lagging edge 
of the secondary electron  peak which corresponds to electrons leaving the sample with $0$
 kinetic energy just outside the surface. The lagging edge  shifts to higher flight time as the
 negative bias is increased demonstrating that PAES is capable of detecting electrons with 
energy more than $0.5$ eV.
} 
\label{fig3} 
\end{figure}
\section{Result and Discussion}
\label{Result and Discussion}
%
%
The PAES spectra from Cu and Au are shown in Fig.~4. The Auger peaks in the case of
  Cu and Au are $M_{23}VV$ and $O_{23}VV$ respectively. The LET can be seen to extend
 from $ 0-50$ eV for Cu and $0-30$ eV for Au.  The argument that the LET is due to the Auger 
electrons only can be ruled out by noting  that secondary electron yield ($\delta$)  
will be $\sim 3.92(3.43)$ for Cu(Au) while the maximum value of $\delta$  for most metals
 is $1.8$ \cite{Joy2005electronYield}. Unlike earlier study of Cu $M_{2,3}VV$ Auger peaks
 \cite{Chiarello1993MVVCu}, we did not observe any plasmon peaks on the low energy side 
of the Auger peak which is consistent with other PAES studies \cite{Zhou1996positron, Hugenschmidt2010PAES}.
\begin{figure} 
\begin{center} 
\includegraphics[width=\hsize,width=8.cm]{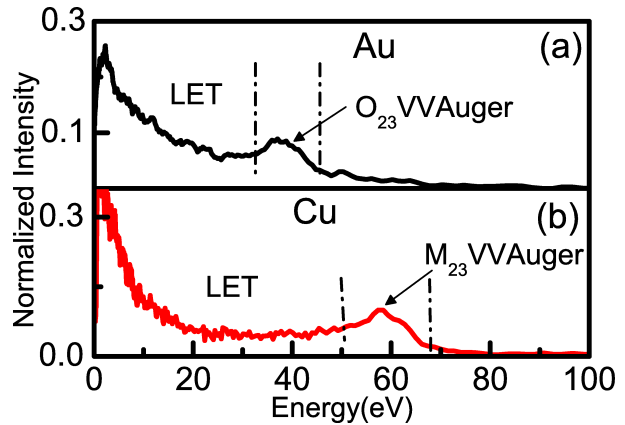} 
\end{center} 
\caption{(Color online)
PAES spectra of (a)Au ($O_{23}VV$) and (b) Cu ($M_{23}VV$ ) obtained using a $1.5$ eV positron beam and a $-0.5$ V bias on the sample.  The  energy scale represents the kinetic energy of the electrons leaving the surface of the sample.  The region enclosed by dashed lines is due to the usual C-VV Augertransition while the  intensity on the left side is the low energy tail.
} 
\label{fig4} 
\end{figure}
%
%

LET associated with the Auger peak has been earlier studied by Auger-photoelectron coincidence
 spectroscopy (APECS) \cite{JensenAPECS, ThurgateIntrinsic} and PAES \cite{Zhou1996positron}. 
Both the studies were limited in the energy range studied ($30-70$ eV for APECS and $18-70$ eV for PAES).
 The APECS  spectra were  further limited as the signal was an average over several atomic layer ($\sim 2$)
 and a  background due to true coincidences involving  inelastically scattered photo electrons from the
 valence band. Earlier PAES experiments employed a positron beam of  
$\sim 18$ eV and hence the low energy part of  such spectra were dominated by extrinsic electrons. In
 the experiments reported here, the Auger signal comes from the topmost layer of the atoms
\cite{ Weiss1988PAES} and the primary beam induced secondary electrons are energetically 
forbidden. Hence the PAES spectrum should contain the major spectral weight in the allowed
 Auger transition energy region, $E_C-2V < E < E_C$, where $E_C$ is binding energy of the core electron,
$V$ is the width of the valence band and E is the energy of the electron. In our experiments, 
we find that majority of the intensity is in LET region (Fig.~4).  
%
%

Following the argument of Ref.~ \onlinecite{JensenAPECS, ThurgateIntrinsic}, the LET 
intensity associated with the Auger peak can be broken down to intrinsic and 
extrinsic contributions 
\begin{equation}
I_{LET}(E)=I_{LET}^{Intrinsic}(E)+I_{LET}^{Extrinsic}(E)
\end{equation}
where $I_{LET}(E)$ is the spectrum in LET region and the factors on the right side of the equation are the intrinsic and extrinsic 
contributions to it. The extrinsic part, in a usual Auger spectrum, is caused by 
primary beam and the transport of the Auger electrons through the solid. 
As suggested in Ref.~\onlinecite{Sickafus1977SecE2, SickafusSecE1}, 
the extrinsic part can be broken down into two components
 \begin{equation}
I_{LET}^{Extrinsic}(E)=I_{beam}(E)+I_{Auger}^{Extrinsic}(E)
\end{equation} 
 where $I_{beam}(E)$ is the primary beam 
induced secondary electron spectrum,  $I_{Auger}^{Extrinsic}(E)$ is the spectrum due 
to the Auger electron scattering inelastically with surface and subsurface 
atoms\cite{Sickafus1977SecE2}. The area under the Auger peak in PAES spectrum
 ($50-70$ eV for Cu and $30-50$ eV for Au) is refered to as $I_{Auger}$. The inherent assumption
 in these analysis is that creation of Auger electrons and their subsequent emission can be
 treated separately \cite{werner2001three}.
 
%
%
As discussed above, we used an incident beam whose energy was below the threshold 
energy that could cause secondary electron emission. It has been demonstrated in Ref.~\onlinecite{Mukherjee2010AMS}
 that the threshold for positron induced secondary electron emission is given by:
\begin{equation}
E_{Kmax}=E_{p}-\varphi^{-}+ E_{b}
\end{equation}
where $E_{Kmax}$ is the maximum energy of emitted electron, 
$\varphi^{-}$ is the electron work function and $E_{b}$ is the binding energy of the positron in 
in the image potential well \cite{Mukherjee2010AMS}. For Cu(Au), 
$\varphi^{-}=4.6(5)$ eV \cite{knights1996effect,*Farjam1987positronPHI}and $E_{b}=2.7(2.9)$ eV and hence the 
threshold for secondary electron emission is $~2$ eV.   The Auger spectra shown in Fig.~4 were 
obtained with an incident beam whose kinetic energy on the sample surface was $1.5$ eV 
thus making the beam induced secondary electron emission channel energetically forbidden. 
%
%

Earlier studies \cite{Mayer1990PAES, *Soininen1991PAES} have shown that backgrounds due 
to positron annihilation-gamma ray induced secondary electrons were not significant in the PAES
 measurements.  These experiments showed that it was possible to turn off the positron annihilation
 induced Auger signal by thermally desorbing the positrons form the surface state into the vacuum
 as positronium Ps (a hydrogen like atom composed of an electron and positron). Most of the Ps
 annihilate in close proximity to the sample.  Consequently if gamma ray induced secondary electrons
were a significant source of background signal, that signal should still be present even after 
the positrons are desorbed from the surface state as Ps. The data in Fig.~5(a) show the results
 of experiments, which are similar to Ref. \onlinecite{Mayer1990PAES, *Soininen1991PAES}. The average
 count of the high temperature spectrum($700^oC$) are only $\sim2\%$ of the RT count rates (obtained 
after cooling the sample to room temperature).  This provides an upper bound on the gamma induced
 secondary background of $\sim2\%$ of the observed signal from the LET and demonstrates that gamma 
induced secondaries are not an important source of background in our measurements.  The fact that
 the signal returned after the sample is cooled demonstrates that heating the sample did not produce
 significant contamination of the surface.
\begin{figure} 
\begin{center} 
\includegraphics[width=\hsize,width=8.cm]{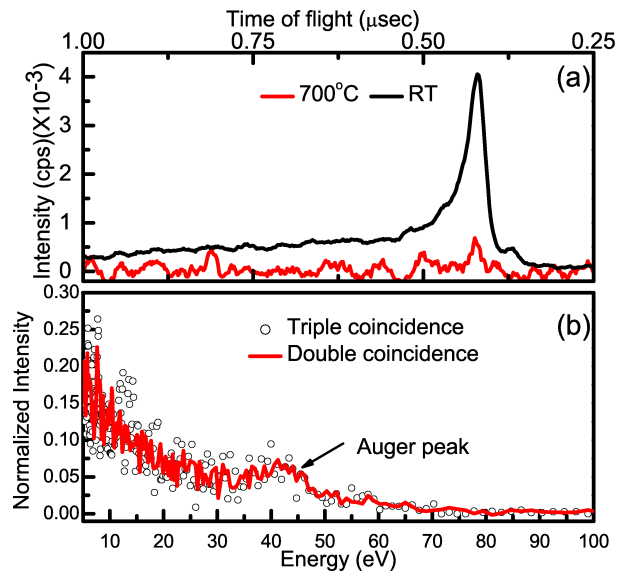} 
\end{center} 
\caption{(Color online)
Estimation of gamma induced secondary electron contribution to LET (a) Comparison of PAES spectrum
 of Cu obtained at room temperature(RT) and at $700^oC$. (b)Normalized PAES and triple coincidence spectra 
from Au surface. The triple coincidence setup is biased against gamma ray induced background.
} 
\label{fig5} 
\end{figure}

 Further verification of the absence of this channel is done by comparing the PAES signal to the
 triple coincidence setup described above. In the latter, a measured coincidence between the two
annihilation gamma rays, emitted at $\sim 180^o$ angle to each other, was required in order to produce
 a valid STOP signal to the TAC. The triple coincidence measurement insured that the annihilation 
events were taking place at the sample since two gamma annihilation taking place at some distance
 from the sample would not be in the simultaneous view of both detectors.  The requirement of
 triple coincidence would strongly suppress background due to annihilation induced secondary electrons
 generated on surfaces other than the sample. The triple coincidence set-up will also strongly suppress
 events where one of the gamma rays undergoes Compton scattering or photo emission in the sample.
  The PAES and the triple coincidence spectrum from Au are shown in Fig.~5(b). The LET region ($0-30$ eV)
 can be seen to have similar intensity in both the PAES and triple coincidence spectra (the poorer statistics
 of the triple coincidence measurements are a consequence of the low count rate associated with the
 reduced joint detection efficiency of the two gamma detectors) proving that the gamma induced
 background has negligible contribution in LET.

%
%

Next, the extrinsic contribution to the LET by the Auger electrons undergoing scattering in the surface
 and subsurface region is explored. The inset in Fig.~6(a) represents an isotropic source of Auger
electrons in the topmost atomic layer \cite{Sickafus1977SecE2}. The electron source is confined to 
the first atomic layer because of the high surface selective nature of PAES. The emitted electrons are 
divided amongst three types based on their probability of undergoing inelastic scattering. Region $1$ 
corresponds to the Auger electrons that escape the solid without suffering any inelastic or elastic
 collision and are eventually detected. These electrons show up as the peak in the Auger spectrum.
  Not all the electrons emitted in the forward direction will fall in this category. The ratio of such 
electrons to all Auger electrons emitted in forward direction is referred to as transmission probability (T).
 Region $2 $ corresponds to those electrons which are emitted at a fairly large angle to the surface normal
and hence will scatter inelastically with the selvedge layer. The ratio of such electrons to the total auger
 electrons in forward direction is given by $1-T$.  As discussed below, these electrons have shown to be 
contributing mostly to the cascade region ($<30$ eV for Cu). Region $3$ corresponds to the Auger electrons
 that are emitted towards the subsurface region and are assumed to be completely lost in the cascade
 process. Since isotropic nature of Auger emission is assumed, the number of electrons in region $3$ will be
 the equal to the sum of those in region $1$ and $2$. The transmission factor, T, has been calculated using 
the Beer lamberts cosine law and is written as  \cite{mehl1990investigation}
 \begin{equation}
T= \frac{\int_0^{2\pi} d\phi \int_0^{\pi/2}exp(\frac{-d}{\lambda_{Cu,Au}cos\theta})sin\theta d\theta}
{\int_0^{2\pi}d\phi\int_0^{\pi/2}sin\theta d\theta}
\end{equation}
 where $\lambda_{Cu,Au}$ is the inelastic mean free path of the Auger electron from Cu(Au) 
\cite{NISTinelastic}, $d$ is the distance the electrons travel in the solid ($1$\AA), $\theta$ is 
the angle from the surface normal and $\phi$ is the azimuthal angle. Hence, T=$50\% (59\%)$ has been 
calculated for Cu-$M_{23}VV$(Au-$O_{23}VV$) Auger electron. The PAES spectra from Cu surface with different
 coverage of residual gases is shown in Fig.~6(b). It can be noticed that the major affect of altering 
the surface roughness or chemistry is in the low energy part of the LET ($<30$ eV for Cu). This leads us to 
 conclude that the Auger electrons from region 2 (Fig.~6(a)) contribute mostly to the cascade region 
of the LET. The secondary electron spectrum produced by inelastic scattering of the Auger electrons 
(represented by 2 and 3) has been modeled as suggested by Ref.~ \onlinecite{Seah1969132} and is 
expressed as $I(E)\sim E(E+E_{PB})^{-1}(E+\Phi)^{-m}$, where $I(E)$ is the intensity of the secondary electron spectrum,
 $E$ is the energy of the electron, $E_{PB}$ is the energy of the primary beam, $\Phi$ is the work function of the
metal and m is a constant. In our case , $E_{PB}$ is taken as the Auger peak energy and is $60$ eV for 
Cu-$M_{23}VV$ and $40$ eV for Au-$O_{23}VV$ transition. This abovementioned line shape is for a beam energy
 of $30$ and $300$ eV normally incident on a polycrystalline copper surface while the electrons emitted 
from one specific angle is detected. This is in contrast to tof-PAES where the spectra are angle
 integrated. Hence, for the sake of comparison, we have measured the $60$ eV positron induced 
secondary electron spectrum from Cu surface using tof-PAES (Fig.~6(a)). Here we have used
 the result that positron and electron induced secondary electron yields tend to be
 similar \cite{ Weiss1989PosSecE}. It can be seen that the shape of the positron induced and
 theoretical secondary electron spectrum are quite similar providing extra confidence in modeling 
the Auger induced secondary electron spectrum by the form mentioned by Ref.~\onlinecite{Seah1969132}. 
The area under
 the secondary electron curve (Fig.6(a)) has been normalized such as to yield a $\delta =0.46(0.21)$
 \cite{Joy2005electronYield} for Cu (Au) with a primary beam intensity of $I_{Auger}[(1-T)/T+1/T]$ .
 Here,  $I_{Auger}$ refers
 to the area under the Auger peak ($50-70$ eV for Cu and $40-50$ eV for Au) while he first term in the
 bracket refers to the electrons from the region $2$ and the last term refers to electrons from 
region $3$. Thus the spectral weight of the extrinsic part can be written as  
$\int^{E_{0}}_0 I^{Extrinsic}_{LET} dE=\delta I_{Auger} [(1-T)/T+1/T]$, where $E_{0}$ represents
the upper limit of LET ($50(30)$ eV for Cu(Au)).

Representing the integrated terms by the respective intensity terms
($\int I(E) dE$$ \rightarrow I$ ) and dividing both sides by the Auger peak area ($I_{Auger}$) , 
\begin{equation}
I^{Intrinsic}_{LET}/I_{Auger}=I_{LET}/I_{Auger}- \delta I_{Auger}[(1-T)/T+1/T]
\end{equation}
Substituting the values Cu(Au) ($I_{LET}/I_{Auger}=3.92(3.43), \delta=0.46(0.21) $ and $T=0.51(0.59)$), the ratio of the intrinsic part of the LET to the Auger peak area ( $I^{Intrinsic}_{LET}/I_{Auger}$) for Cu(Au) is $2.4(3.0)$. 

\begin{figure} 
\begin{center} 
\includegraphics[width=\hsize,width=8.cm]{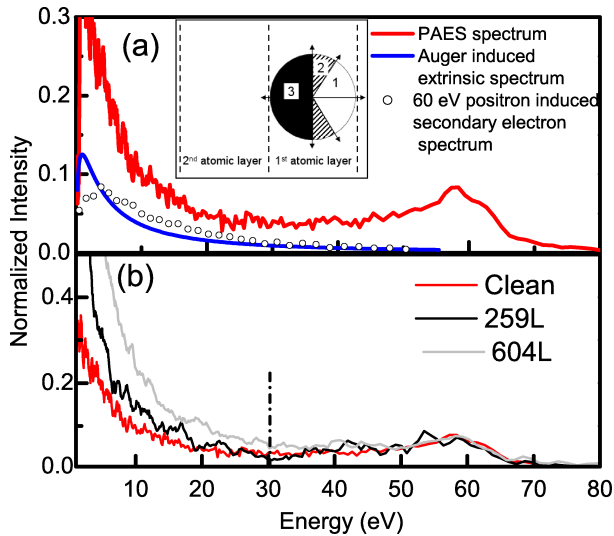} 
\end{center} 
\caption{(Color online)
Estimate of the extrinsic background due to the Auger electrons scattering in the surface and
 subsurface region from Cu surface. (a)The normalized secondary electron spectrum as suggested
 by Ref \onlinecite{Seah1969132} has been compared to the $60$ eV positron induced secondary electron
 spectrum as recorded by TOF-PAES. The inset shows an isotropic Auger source located in the 
topmost atomic layer[Region1=the Auger electrons which are emitted without any inelastic scattering,
 Region $2$=Auger electrons which scatter inelastically from the surface, Region $3$=Auger electron 
scattering from the subsurface region \cite{Sickafus1977SecE2}] (b) Normalized Cu-PAES spectra 
with different coverage of residual gas. As can be seen, the effect of surface scattering  is mostly
 in the cascade region ($<30$ eV).
} 
\label{fig6} 
\end{figure}
%

%
%

The Cu-$ M_{23}VV$  and Au-$O_{23}VV$ Auger transition with the estimated extrinsic contribution
subtracted is shown in Fig.~7. The ratio of  spectral weight  in the main Auger peak to 
the intrinsic LET region is $1\colon2.6(3.1)$. The intrinsic  LET can be interpreted as due to 
the core holes decaying via  multi electron emission process and is referred to as C-VVV process\cite{ramaker:1614}.
The intrinsic emission is mostly thought of in terms of  sudden limit 
which means that the core hole was created in a timeframe less than that of the
Auger lifetime ($\sim fs$) \cite{HimpselAdiabaticAuger1980}. If the time taken to create the core hole
is less then the Auger lifetime then the process is termed adiabatic \cite{Darrahadiabatic1984}.
Adiabatic process is evident mostly in resonant emission as the outgoing photoelectron has 
small velocity so that the valence electrons, effectively, see no change in the effective charge.
To establish the timescale in which the core hole is created in PAES, we have to consider the
time taken by the positron to tunnel into the core region ($\sim100$ ps) \cite {FertigTunnelingtime1990, *SteinbergTunnelingtime1990, *ChuprikovTunneling2006-1, *ChuprikovTunneling2006-2, Mills}. Hence the valence electrons see some
effective charge, say Z, which changes slowly ($\sim100 ps$) from $Z+1$ as the positron tunnels
 into the core region. This is followed by rapid annihilation ($\sim10^{-23}$ sec)\cite{Mills} while the
effective charge seen by the valence electrons remains at $Z+1$. Hence the PAES process
 can be considered to be in adiabatic regime and here we show that such emissions
 are possible in adiabatic regime as well.
\begin{figure} 
\begin{center} 
\includegraphics[width=\hsize,width=8.cm]{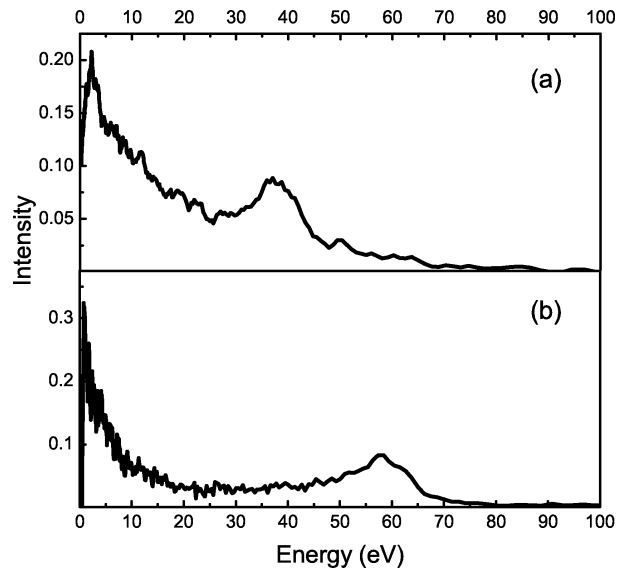} 
\end{center} 
\caption{(Color online)
CVV Auger spectra of (a) Au and (b) Cu after the subtraction of extrinsic contributioon. The intrinsic  LET (C-VVV transition) extends  to $\sim0$ eV and  is $\sim~2.4 (3.0)$ times as intense as the Auger peak (C-VV transition) in case of Cu(Au).
}
\label{fig8} 
\end{figure}

The spectrum in Fig.~7 has been used to calculate the probability of core holes decaying 
via C-VVV process.  Since such a process entails emission of two electrons (sharing the energy of the usual Auger electron), 
the spectral weight of the C-VVV process will be twice that of the conventional CVV process.
 Hence the percentage of core holes decaying via multi electron emission for Cu(Au) has been 
calculated to be $\sim54(60)\%$. The C-VVV emission probability has also been used to
 calculate the positron-core electron annihilation probability at surfaces. Earlier
 calculations \cite{PhysRevB.41.3928} considered only the the core-holes which
 decayed via the CVV process. Hence a $3.1\%$ and $3\%$ core annihilation
 probability was calculated for Cu($3p$) and Au($5p$).  Based on our experiments, we 
find the respective annihilation probability would  be $6.8\%$ and $7.5\%$ which compares
 well with recent theoretical estimate \cite{fazleev2010oxidation, AlataloCoreannProb1996}.
\section{Conclusion}
\label{Conclusion}
We have reported the first background free measurement of the complete spectra 
 of low energy electrons 
   emitted as a result of Auger transition in metals. By depositing low energy ($\sim 1.5 eV$) positrons
 directly into the surface state \cite{ Mukherjee2010AMS}it was possible to excite Auger
  transitions from atoms at the surface without generating any primary beam induced secondary
 electrons. Localization of the positrons in the surface state ensures that almost all of Auger
 transitions occur in the topmost atomic layer.   The resultant spectra showed that majority of
 the spectral weight is in the low energy tail associated with the Auger peak and extends 
to $0$ eV.
Estimates based upon measurements of the electron induced secondary electron yield 
and inelastic scattering probability of electrons with selvedge indicate that extrinsic contributions
 account for less than $36(14)\%$ of the LET in case of Cu(Au). Our results suggest that the
 intrinsic part of  LET is due to the process in which  the core hole decays by emission of more
 than one Auger electron (C-VVV). Assuming that the intrinsic process involves emission of only two
 electrons, it has been calculated that $54 (48)\%$ of the core holes in Cu(Au) decay via multi
 electron emission. This result was used to obtain new estimate of positron-core electron 
annihilation probability for Cu and Au which agree well with theoretical calculations. Since in
 the case of multi electron Auger emission, the valence electrons are emitted simultaneously,
 our studies are analogous to spectroscopy of photon induced emission of electron pairs
 \cite{PhysRevLett.89.086402} and provide another way of probing  electron correlation
  effects in valence band of metals.
Our results also have important implications in quantitative analysis of Auger 
spectra\cite{seah1996quantitative}.  In particular, we show that a significant
 fraction of core holes decay via multiple electron processes.  These processes
 result in a decrease in the Auger peak intensity that is not fully accounted for
 in the usual Beer-Lambert based calculations of the escape probability.
  Consequently, estimates based upon measuring of the integrated intensity 
 in the Auger peak region alone may lead to an underestimate of the number
 of initial core hole excitations.  This has important implication for the use of PAES
 in estimating core hole annihilation probabilities \cite{fazleev2010oxidation} and
 for the use of Auger spectroscopy in the quantitative analysis of surfaces.      
\section{Acknowledgement}
\label{Acknowledgement}
 We wish to acknowledge useful discussion with D. E. Ramaker, J.Moxom and A.G.Hathaway. This work was
 supported by the Welch Foundation, Y1100 and  NSF Grant No.$DMR-0907679$.

\bibliography{mybib}

\end{document}